# Electrical Stark Tuning and In Situ Field Sensing of $V_2$ Silicon Vacancies in a 4H-SiC a-Plane Lateral pin-Diode

**Jannik H. Schwarberg[1,*], Fedor D. Hrunski[2], Maximilian Hollendonner[2], Wolfgang Knolle[3], Mathias Rommel[4], Fabian J. Magerl[1], Jan F. Dick[1,4] and Jörg Schulze[1,4]**

[1] Chair of Electron Devices, Friedrich-Alexander-Universität Erlangen-Nürnberg, 91058 Erlangen, Germany
[2] Institute of Applied Quantum Technologies, Friedrich-Alexander-Universität Erlangen-Nürnberg, 91058 Erlangen, Germany
[3] Leibniz-Institut für Oberflächenmodifizierung, 04318 Leipzig, Germany
[4] Fraunhofer Institute for Integrated Systems and Device Technology IISB, 91058 Erlangen, Germany

*Author to whom any correspondence should be addressed.

**E-mail:** jannik.schwarberg@fau.de

## Abstract

Silicon carbide has emerged as a highly promising platform for scalable quantum technologies, with the silicon vacancy serving as a robust spin qubit and quantum sensor. However, conventional c-plane 4H-SiC restricts the experimental scheme and resonant optical excitation to the sample's geometry. The usage of a-plane substrates enables more versatile resonant optical access to the color center through the wafer surface. Here, we demonstrate the electrical control and optical addressability of V2 (cubic lattice site) silicon vacancies within the intrinsic region of lateral pin-diodes fabricated on a-plane 4H-SiC wafers using a scalable complementary metal-oxide-semiconductor (CMOS) process. Despite extensive semiconductor processing, the integrated color centers exhibit good optical quality, evidenced by narrow optical transition linewidths below 80 MHz at zero bias. These embedded quantum sensors are utilized for in situ measurement of local electric fields at cryogenic temperatures via the Stark effect. Comparisons with macroscopic technology computer-aided design simulations reveal the value of these integrated quantum sensors to help understand cryogenic device behavior, for example in predicting highly localized field crowding. Furthermore, to address the spectral indistinguishability required for quantum networks, we precisely tune the zero-phonon line (ZPL) of individual emitters via the reverse-bias Stark effect. This successfully compensates for the spectral variations of distinct emitters. Following the relaxation of transient charge fluctuations, the Stark-tuned emitters exhibit spectral stability over measurement periods exceeding four hours. Our results establish a-plane 4H-SiC as a robust platform for future interference experiments and quantum photonic integrated circuits.

## 1. Introduction

Silicon carbide (SiC), particularly the 4H polytype, has emerged as a highly promising platform for quantum technologies due to its ability to host deep-level color centers, such as the silicon vacancy ($V_{Si}$) [1–6], while offering a mature, scalable semiconductor manufacturing ecosystem [7, 8]. The $V_{Si}$ acts as a localized spin qubit with robust spin-optical properties, making it an excellent candidate for applications ranging from high-precision quantum sensing to scalable quantum computing using quantum photonic integrated circuits (QPICs) [1, 9–11]. However, transitioning these demonstrations from isolated laboratory setups into robust, scalable, and industrially relevant technologies requires overcoming significant material and device engineering challenges.

A critical parameter for the optical and electrical integration of $V_{Si}$ is the crystallographic orientation of the substrate. Typically, c-plane 4H-SiC substrates are used for semiconductor device manufacturing. In these conventional substrates, the dipole moment of the $V_{Si}$ is oriented perpendicular to the wafer surface (neglecting the 4° off-cut needed for epitaxial layer growth) [12]. This geometry necessitates resonant laser excitation from the cleaved edge of the sample [2] or complex photonic structures [13–16], which limits investigations to single vertical devices and complicates planar integration. In contrast, the use of a-plane 4H-SiC substrates aligns the c-axis – and therefore the $V_{Si}$ dipole moment – parallel to the surface [17]. This fundamental shift enables direct, out-of-plane resonant optical excitation, paving the way for the fabrication of lateral device architectures using standard scalable semiconductor processing [18, 19]. Moreover, this in-plane dipole orientation natively matches the horizontal electric field vector of transverse-electric (TE) photonic waveguide modes, enabling direct and efficient integration into standard TE-polarized planar circuits [20, 21].

By integrating $V_{Si}$ into lateral pin-diodes on a-plane wafers, two primary application domains can be addressed. First, in the context of device optimization, $V_{Si}$ can act as embedded *in situ* quantum sensors to map local electric fields within operating power electronic devices [2, 22, 23]. This provides three-dimensional internal feedback that is currently impossible to obtain via external sensors, allowing for the precise calibration of technology computer-aided design (TCAD) simulations and the identification of localized failure mechanisms, such as field crowding [24].

Second, for quantum computing and integrated quantum networks, $V_{Si}$ can serve as individual qubits [3, 25]. A fundamental requirement for multi-qubit entanglement is the generation of indistinguishable photons [12, 26]. Due to microscopic variations in their local environment, different $V_{Si}$ naturally emit at slightly different zero-phonon line (ZPL) frequencies [12]. The lateral pin-diode architecture allows for the application of highly controlled, directional electric fields parallel to the c-axis, enabling the precise Stark tuning of the emission frequency of single $V_{Si}$ centers [2, 27]. Consequently, distinct emitters can be shifted into (or out of) spectral resonance, fulfilling a core prerequisite for generating indistinguishable photons within a scalable CMOS architecture.

## 2. Methods

Lateral pin-diodes were fabricated on 4H-SiC a-plane substrates featuring a low n-type doped epitaxial layer using a scalable CMOS process [7, 18, 28]. Highly doped p-type and n-type regions were formed via aluminum and nitrogen ion implantation followed by high-temperature thermal activation and contacted with silicided ohmic contacts and metallization stacks. To enable low-noise optical access to the intrinsic region, a passivated optical window preserving a high quality 50 nm thermal oxide was structured above the active area using a protective poly-Si layer and selective wet etching [29]. Subsequently, silicon vacancy ($V_{Si}$) color centers were generated via electron irradiation and thermal annealing, remaining optically active exclusively within the low-doped intrinsic region due to charge-state stability [30, 31]. The final device is shown in Figure 1. See Supplementary Information Section SI 1 for fabrication details.

To evaluate the internal electrostatics and electric field profiles at 300 K, two-dimensional numerical simulations of the fabricated diode geometry as shown in Figure 2 were conducted in Synopsys Sentaurus TCAD. The model incorporates crystallographic anisotropy, incomplete dopant ionization, and an imperfect p-type Schottky-tunneling contact [18] calibrated against experimental forward *I-V* characteristics via numerical optimization (detailed in Supplementary Information Section SI 2).

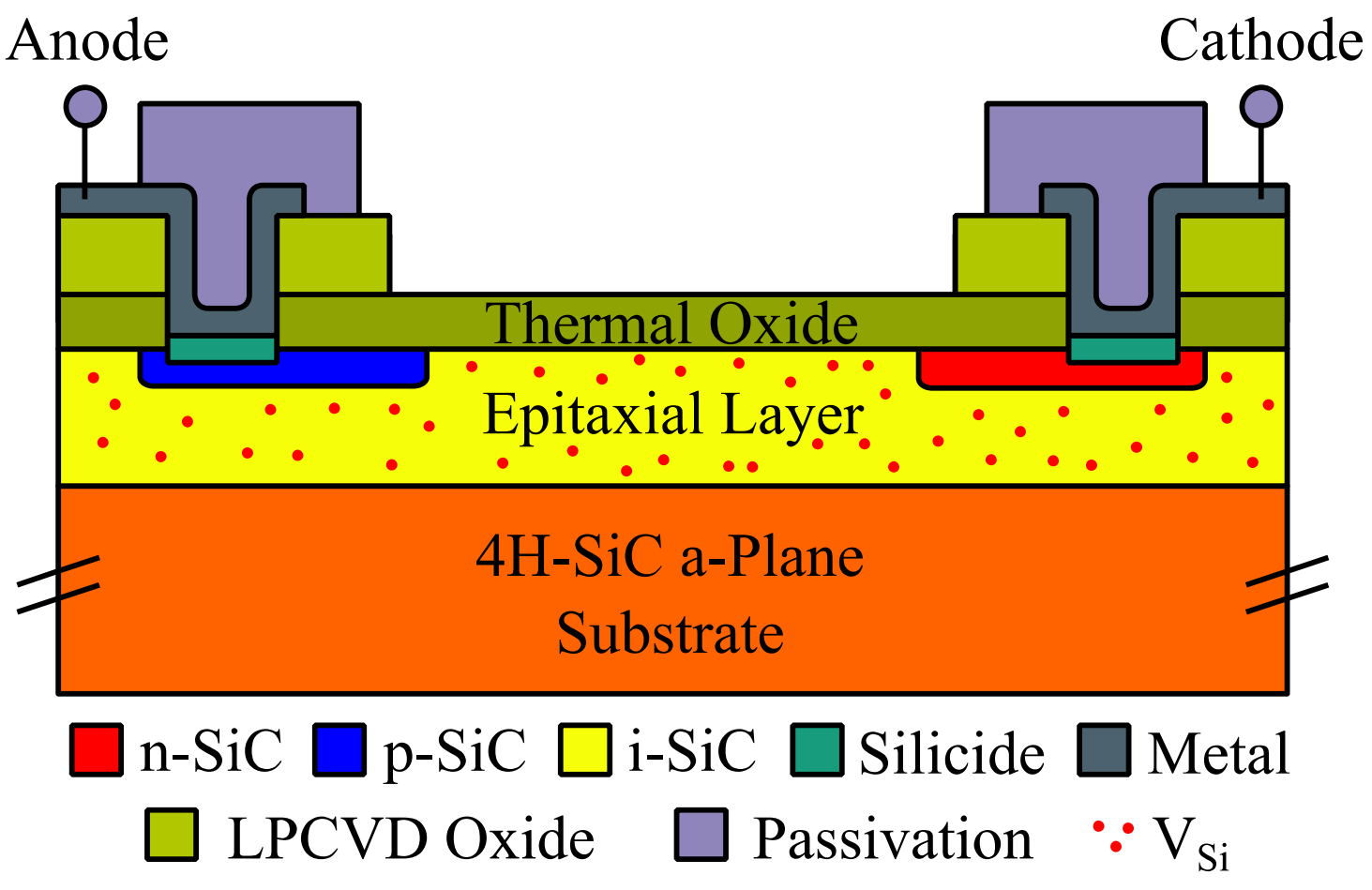

**Figure 1.** Schematic cross-section of the fabricated lateral pin-diode on a 4H-SiC a-plane substrate, illustrating the structured optical window and the generated silicon vacancies within the intrinsic region.

Cryogenic optical characterization of the $V_2$ centers was performed at 4 K using a home-built confocal microscope [32]. Color centers were excited off-resonantly at 728 nm and resonantly at 916 nm (including double-tone PLE modulation), while filtered phonon-sideband emission (>950 nm) was detected using superconducting nanowire single-photon detectors (see Supplementary Information Section SI 3 for the full experimental setup).

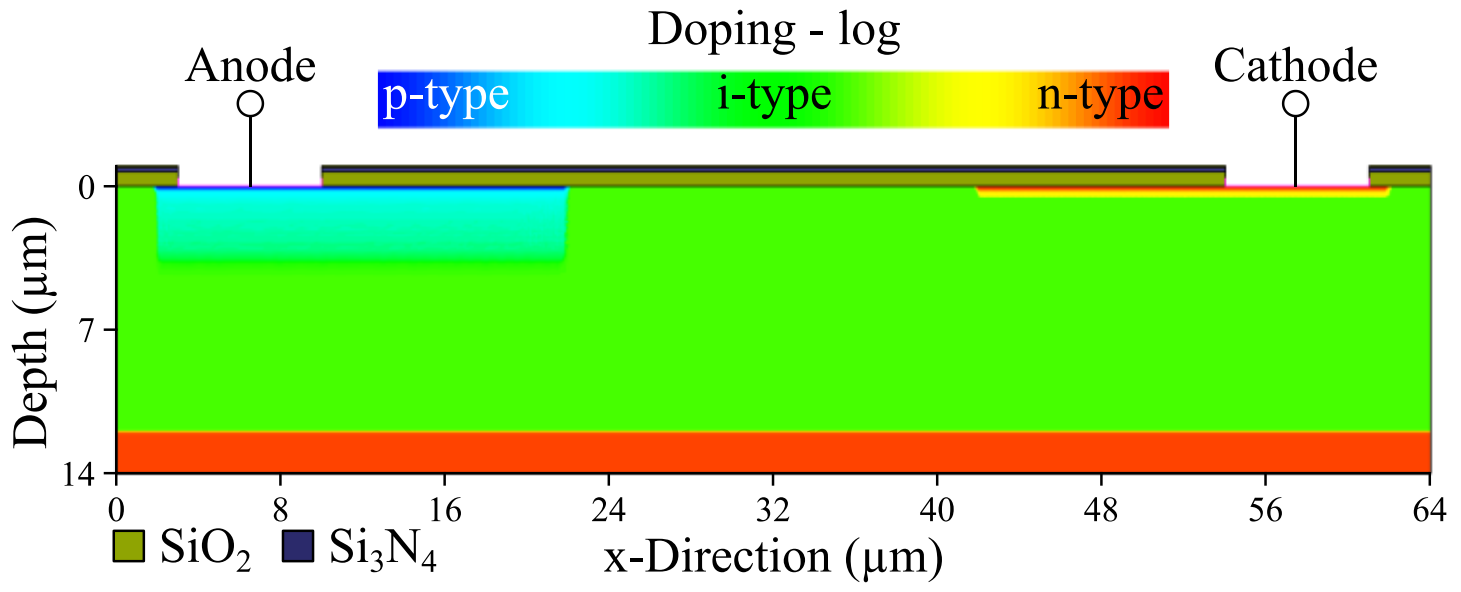

**Figure 2.** Two-dimensional cross-sectional doping profile (logarithmic scale) of the lateral pin-diode simulated in Sentaurus TCAD on an a-plane 4H-SiC substrate with a lowly n-type doped ($N_D = 2.4 \times 10^{13}$ cm$^{-3}$) epitaxial layer. The x-direction of the device is aligned parallel to the crystallographic c-axis.

## 3. Results

### 3.1 Electrical Characterization and Calibration of TCAD Simulations

The fabricated lateral pin-diodes show good blocking capabilities up to −150 V and reverse leakage currents below the noise floor of our measurement setup (<10 pA) [29], which is essential for minimizing charge noise during optical measurements. A comprehensive statistical analysis of these electrical characteristics is detailed in Ref. [19]. To accurately correlate the applied external voltages with the local electric fields interacting with the color centers, two-dimensional TCAD simulations were conducted and calibrated against the experimental data.

Because measurement limits made it impossible to determine the extremely low reverse current exactly, the physical models were calibrated using the forward *I-V* characteristics (Figure 3 – black dots). Initial simulations utilizing standard parameters from Sentaurus TCAD for 4H-SiC exhibited a significant discrepancy compared to the experimental forward current, failing to reproduce the measured turn-on behavior and series resistance, as can be seen in Figure 3 (red dashed line). The calibrated parameters, as detailed in SI 2, were determined as follows: an effective carrier lifetime of 6.78 ns, a barrier height of 1.01 eV, an effective tunneling mass of $0.16 \times m_0$, a parasitic series resistance of 16.17 Ω cm$^2$, and an effective width of 27.68 μm (design width is 20 μm). By applying these optimized parameters, the calibrated simulation (green dashed line) achieves high agreement with the experimental data, as shown in Figure 3.

To understand the electrostatics during the optical manipulation of the silicon vacancies, the calibrated models were adapted to simulate the reverse-bias regime. Because the p-contact acts merely as a parasitic barrier under reverse bias rather than a tunneling contact, the tunneling models were disabled, and high-field extensions such as anisotropic avalanche and field-dependent SRH recombination were activated. Under a reverse bias of −50 V, the internal potential drops continuously across the intrinsic epitaxial layer. The simulations reveal a pronounced local enhancement of the electric field (field crowding [24]) at the corner of the p-type implanted region facing the intrinsic zone (see Figure 4 – left partial cross section), caused by the extreme spatial convergence of the electric field lines.

Unlike in vertical pin-diodes, this lateral geometry creates a highly complex position-dependent electric field. Near the p-type region, the electric field vector possesses a significant component along the crystallographic a-axis. For instance, for −50 V applied voltage at $x = 22$ μm in a depth of 200 nm it is tilted by 27.2° towards the surface. In contrast, in the central region between the contacts near the oxide interface, the electric field lines run predominantly parallel to the semiconductor surface, aligning with the c-axis (see Figure 4 – right partial cross section). Here, the tilt at $x = 32$ µm in a depth of 200 nm is below 5° for the same applied voltage. The simulations of this device revealed that independent of position, the magnitude of the electric field rises almost linearly for reverse bias voltages below −5 V, but with a position dependent gradient. Thus, when considering the Stark shift, a potential small built-in electric field can be neglected and the gradient of the electric field and the corresponding PLE shift can be discussed.

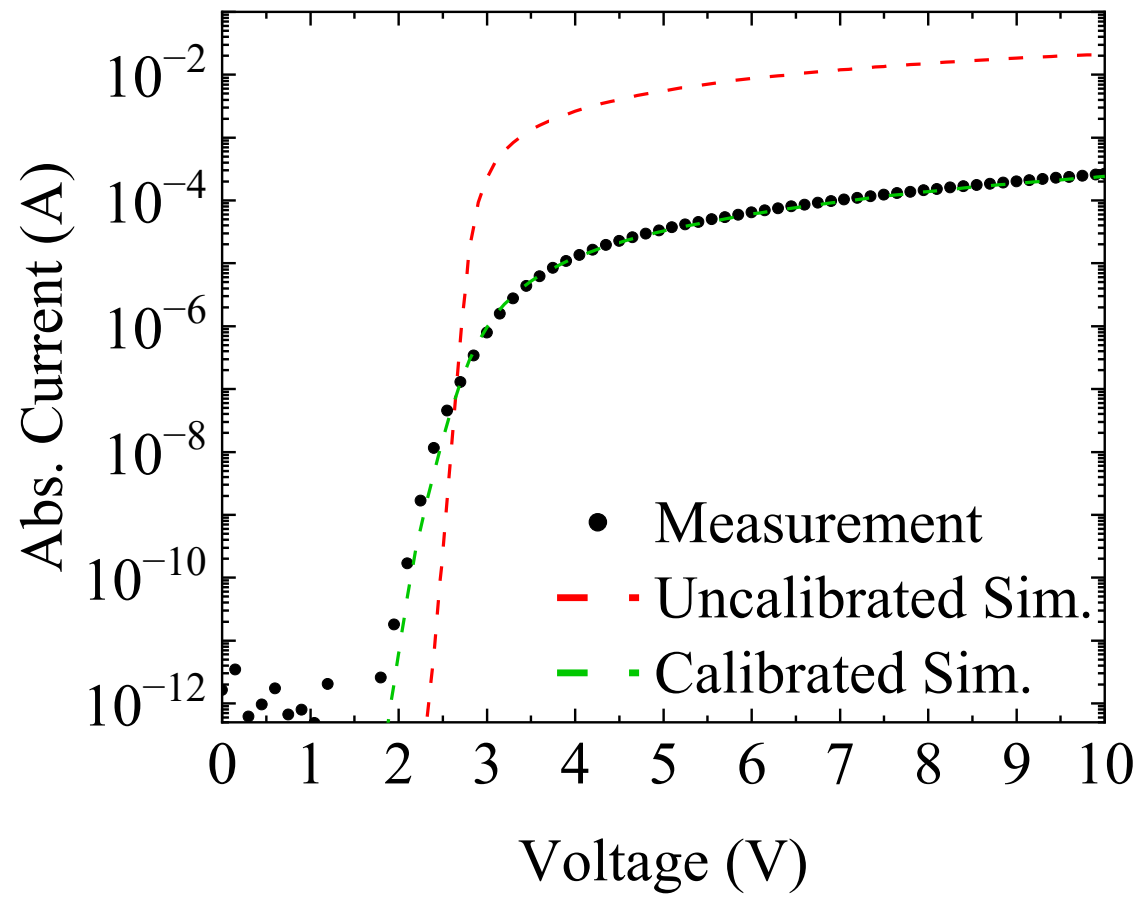

**Figure 3.** Experimental and simulated forward *I-V* characteristics of the lateral pin-diode. The uncalibrated simulation using standard 4H-SiC parameters shows a significant deviation, whereas the calibrated simulation accurately reproduces the measured *I-V* characteristic.

This precise spatial mapping of the electric field vector is critical, as the Stark effect relies strongly on the relative orientation of the local electric field to the defect's dipole moment [2, 23, 33]. Furthermore, the simulations confirm that without applied voltage (not shown) as well as at −50 V, the entire relevant intrinsic region is depleted of mobile charge carriers due to the low doping of the epitaxial layer.

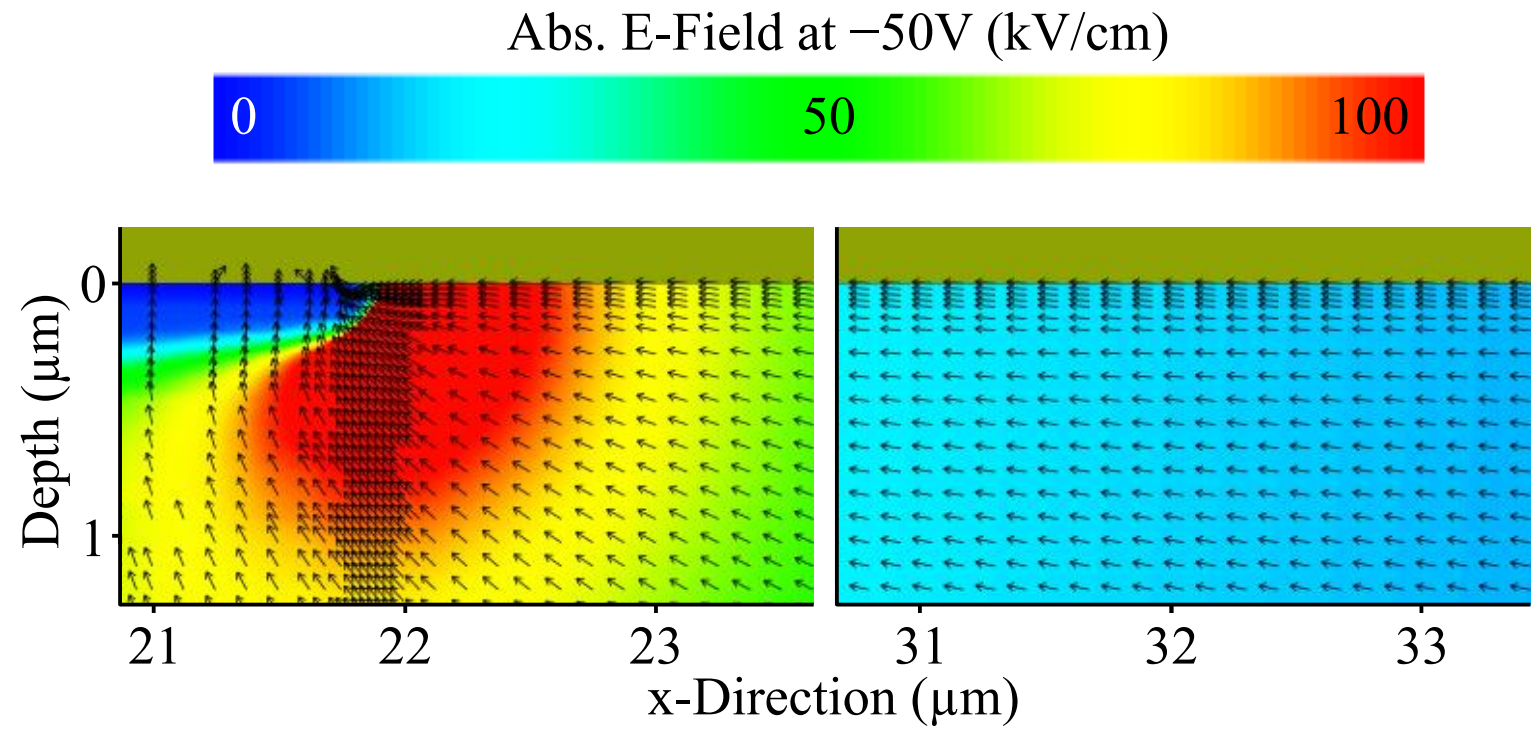

**Figure 4.** Spatial distribution of the absolute electric field within the intrinsic region of the lateral pin-diode (excerpts from Figure 2) under a reverse bias of −50 V at 300 K, highlighting the local field enhancement near the p-type region (left part) and the uniform, surface-parallel field alignment in the central region (right part).

### 3.2 Stark Effect on Silicon Vacancies in the pin-Diode

Following the electrical characterization of the devices, the integrated $V_2$ silicon vacancies within the active intrinsic region of the pin-diode were localized using confocal photoluminescence microscopy. Figure 5 shows an overview scan where the strong background luminescence identifies the p-type region on the left, while most of the bright spots in the central region correspond to color centers. The differences in brightness among the emitters indicate slightly varying depth positions within the epitaxial layer.

The exact frequencies of the spin-preserving transitions $A_1$ (between $|\pm 1/2\rangle$ states) and $A_2$ (between $|\pm 3/2\rangle$ states) for these centers were first determined under zero external bias. An

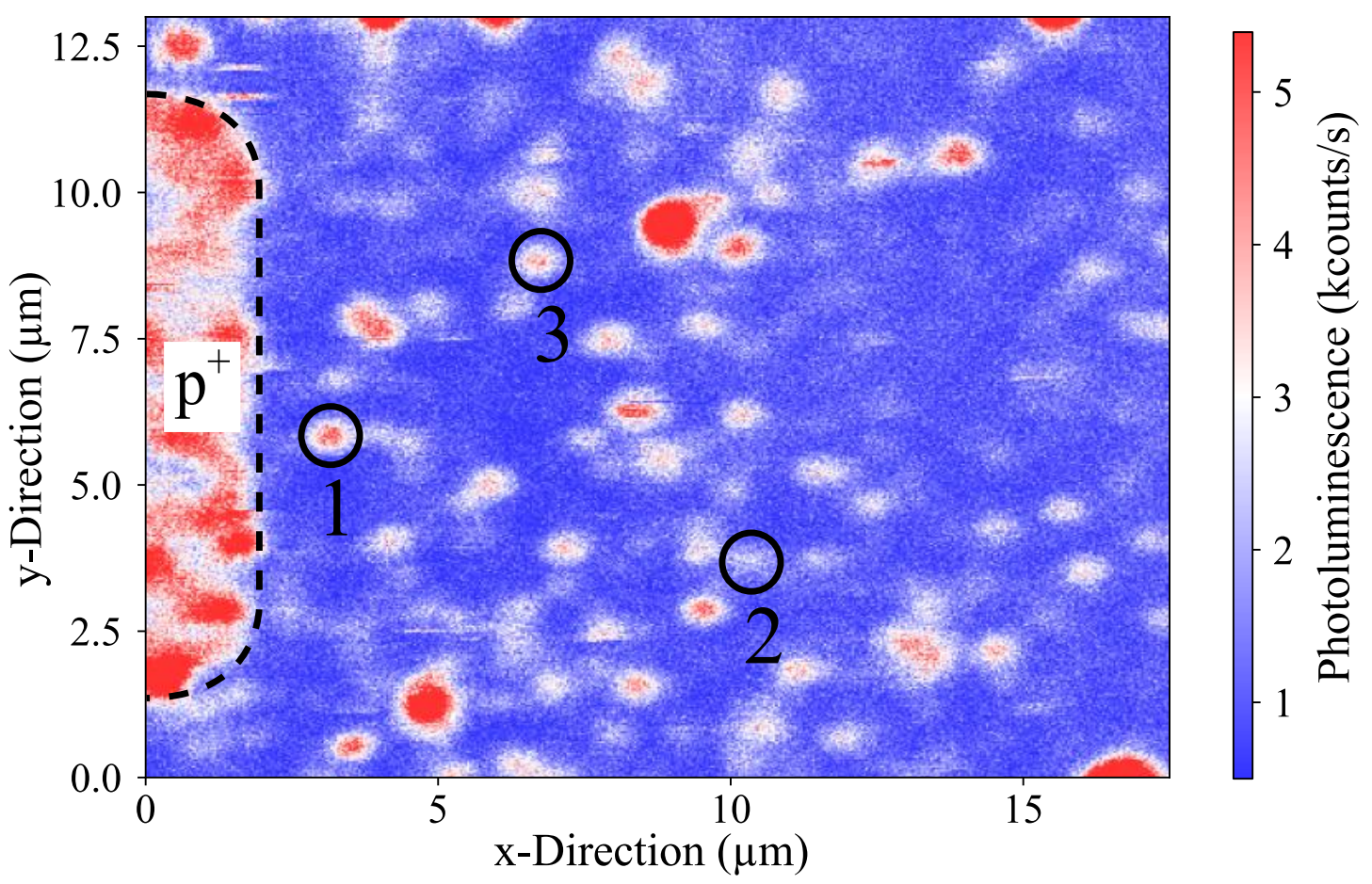

**Figure 5.** Confocal PL scan of the active diode region at 4 K. The p-type region is visible on the left due to strong background luminescence. The numbered bright spots mark the investigated $V_2$ silicon vacancies (No. 1 to No. 3) located within the intrinsic region.

exemplary measurement is shown in the inset in Figure 6a. Across six investigated emitters, the standard deviation of the zero-bias $A_1$ resonance frequency was 23.9 GHz (which is comparable to the findings in Ref. [12]), pointing out the necessity to tune these emitters to enable deterministic spectral matching.

To precisely determine the optical transition linewidths without microwave-induced power broadening, all-optical photoluminescence excitation (PLE) spectroscopy was performed using sideband modulation via an electro-optic modulator (EOM). By fixing the laser to the $A_1$ transition and scanning the EOM modulation frequency to drive the $A_2$ transition, the spectral linewidth could be measured precisely. Using this method at 0 V bias, the FWHM of the optical $A_1$ and $A_2$ transitions of color center No. 2 were found to be 69.4 MHz and 75.6 MHz, respectively.

To overcome the spectral variations between individual emitters and tune them to identical emission wavelengths, the Stark effect induced by the reverse bias of the pin-diode was utilized. The Stark shift was analyzed for color centers No. 1 (Figure 6a) and No. 2 (Figure 6b), demonstrating significantly different responses depending on their position within the diode. Color center No. 1, located near the junction from p-type to the intrinsic region (which is slightly n-type doped) where the electric field lines converge (consistent with the TCAD simulation of field crowding shown in Figure 4 – left part), exhibited a strong Stark shift of 4.79 GHz/V. In contrast, center No. 2 is located further into the intrinsic region where the electric field and its gradient are weaker, resulting in a shift of approximately 0.32 GHz/V.

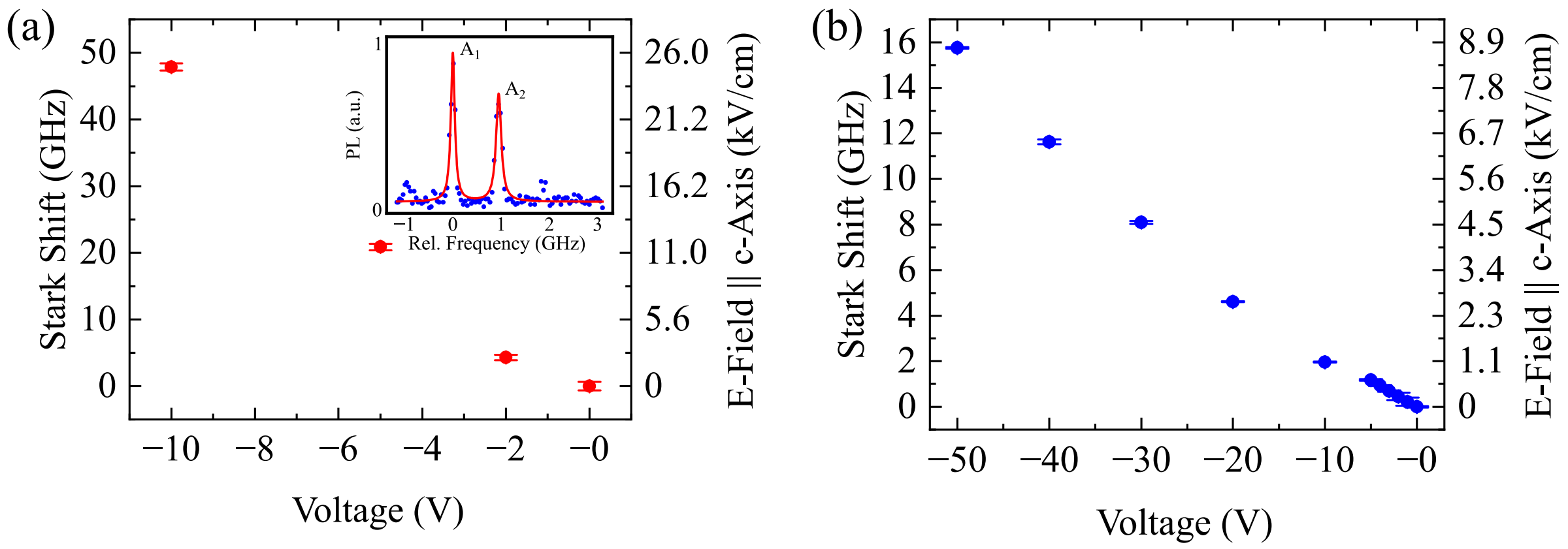

**Figure 6.** Voltage-dependent Stark shift of the resonance frequency for color center No. 1 (a) and No. 2 (b) from Figure 5. The right y-axis features a non-linear scaling that translates the measured frequency shift into the corresponding absolute electric field parallel to the c-axis, calculated using the quadratic Stark effect model. The inset in Figure 6a displays an exemplary all-optical PLE spectrum used to precisely determine the $A_1$ and $A_2$ transitions with a double Lorentzian fit.

The observed frequency shifts can be correlated with the local electric field, because the optical shift of the ZPL is primarily governed by the electric field component parallel to the c-axis $E_{\parallel}$ [2, 27, 33]. This relationship is given by solving the equation from Ref. [27] for $E_{\parallel}$:

$$E_{\parallel} = \frac{-d_{\parallel} + \sqrt{d_{\parallel}^2 - 2 \cdot \alpha_{\parallel} \cdot \Delta f_{\mathrm{PLE}}}}{\alpha_{\parallel}} \tag{1}$$

where $\Delta f_{\mathrm{PLE}}$ is the Stark shift of the resonance frequency, $d_{\parallel}$ is the dipole moment, and $\alpha_{\parallel}$ is the polarizability along the c-axis. For this work the electric field is the macroscopic field, commonly used in semiconductor physics without consideration of the Lorentz local field approximation [34] (see SI 4 for details). The parameters used for the calculation of the electric field are from Ref. [2] (chosen because both studies employed the same measurement procedure) adapted for using the macroscopic field ($d_{\parallel} = 17.32\ \mathrm{GHz/(MV/m)}$ and $\alpha_{\parallel} = -1.44\ \mathrm{GHz/(MV/m)^2}$). The right axis in the corresponding graphs in Figure 6 is scaled non-linearly according to Equation 1 to reflect the quadratic term in the Stark equation, directly translating the frequency shift to the absolute electric field in the c-direction. For a bias voltage of −10 V the increase of the electric field was determined to be 25.0 ± 0.7 kV/cm for color center No. 1 and 1.13 ± 0.05 kV/cm for color center No. 2.

To complement the purely optical investigations, the effects of the applied reverse bias on the ground state spin properties were characterized. To overcome the low signal contrast and long integration times typical for off-resonant optically detected magnetic resonance (ODMR) measurements [2, 35–37], resonant ODMR was performed at 4 K on color center No. 2 for various applied reverse bias voltages.

By continuously adjusting the excitation laser to the exact Stark-shifted $A_1$ transition frequency for each applied voltage step, the ODMR signal contrast was significantly enhanced. An exemplary raw measurement (at −25 V) of this resonant ODMR (Figure 7a) exhibits a pronounced resonance peak at 68.37 ± 0.06 MHz. This technique enables the simultaneous extraction of both the optical PLE shift and the ODMR ground state spin resonance frequency, ensuring both measurements reflect identical internal electric field conditions. Over a reverse bias range up to −30 V (see Figure 7b), the voltage-dependent ODMR resonance frequency was measured to decrease from approximately 68.62 ± 0.05 MHz at −3 V down to 68.30 ± 0.05 MHz at −30 V (left axis). This correlates with the optical Stark shift affecting the $A_1$ transition of the excitation wavelength (right axis).

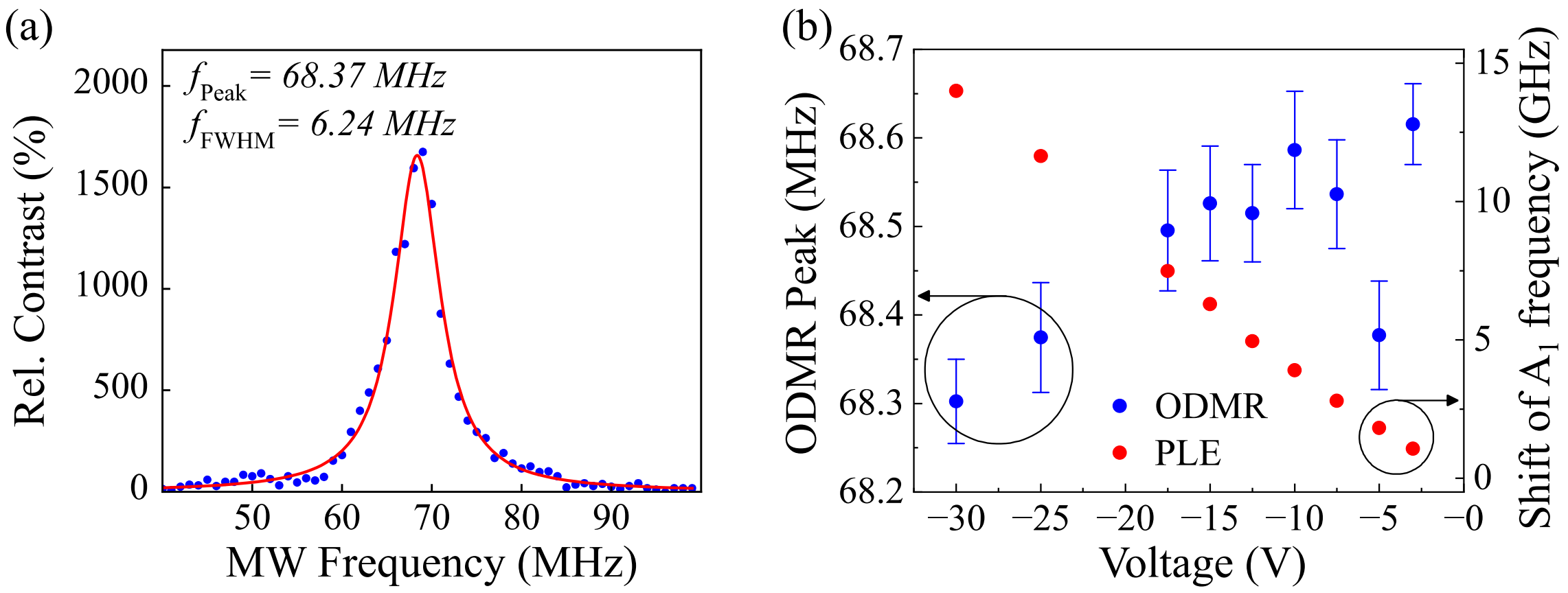

**Figure 7.** Resonant ODMR measurements of color center No. 2 at 4 K. (a) Exemplary high-contrast raw ODMR spectrum showing a resonance peak at 68.37 MHz. (b) Correlation between the voltage-dependent ground state ODMR resonance frequency (left axis) and the shift of the $A_1$ frequency between ground and excited state used for resonant excitation during these resonant ODMR measurements (right axis) for varying reverse bias voltages. The black circles indicate which axis the plotted values belong to.

### 3.3 Transient Influences on the Spectral Stability of the Resonance Frequency

Time-resolved PLE scan series were performed to monitor the stability of the ZPL. The exact spectral position of the resonance frequencies exhibits sensitivity to transient perturbations caused by both optical and electrical sources.

First, the influence of the off-resonant initialization laser was evaluated, as shown in Figure 8a and 8c. The measurement was intentionally interrupted for extended durations (~ 1 min) between individual scans to expose the color center to the 728 nm laser (~ 470 μW). When the measurement was subsequently continued under the exact same external conditions as before the pause, a sudden, temporary shift of the optical transitions toward lower resonance frequencies was observed. Following this induced shift, the resonance frequency gradually drifts back to its original equilibrium position over the course of several minutes.

Second, the transient stability of the resonance frequency was evaluated following abrupt changes in the applied reverse bias, establishing a dependence on the device's previously applied voltages. In a measurement series, shown in Figure 8b and 8d, performed at a constant target bias of −40 V, the scan was paused temporarily (< 1 min) to apply a different voltage, and then resumed at the original −40 V target bias. Stepping the device from a larger reverse bias (e.g., −50 V) down to the target −40 V results in a short-term shift to higher frequencies, which slowly relaxes to the equilibrium state. Conversely, stepping the device from a smaller reverse bias (e.g., −30 V or −20 V) up to −40 V causes a transient shift to lower frequencies, which subsequently rises back to the steady-state equilibrium.

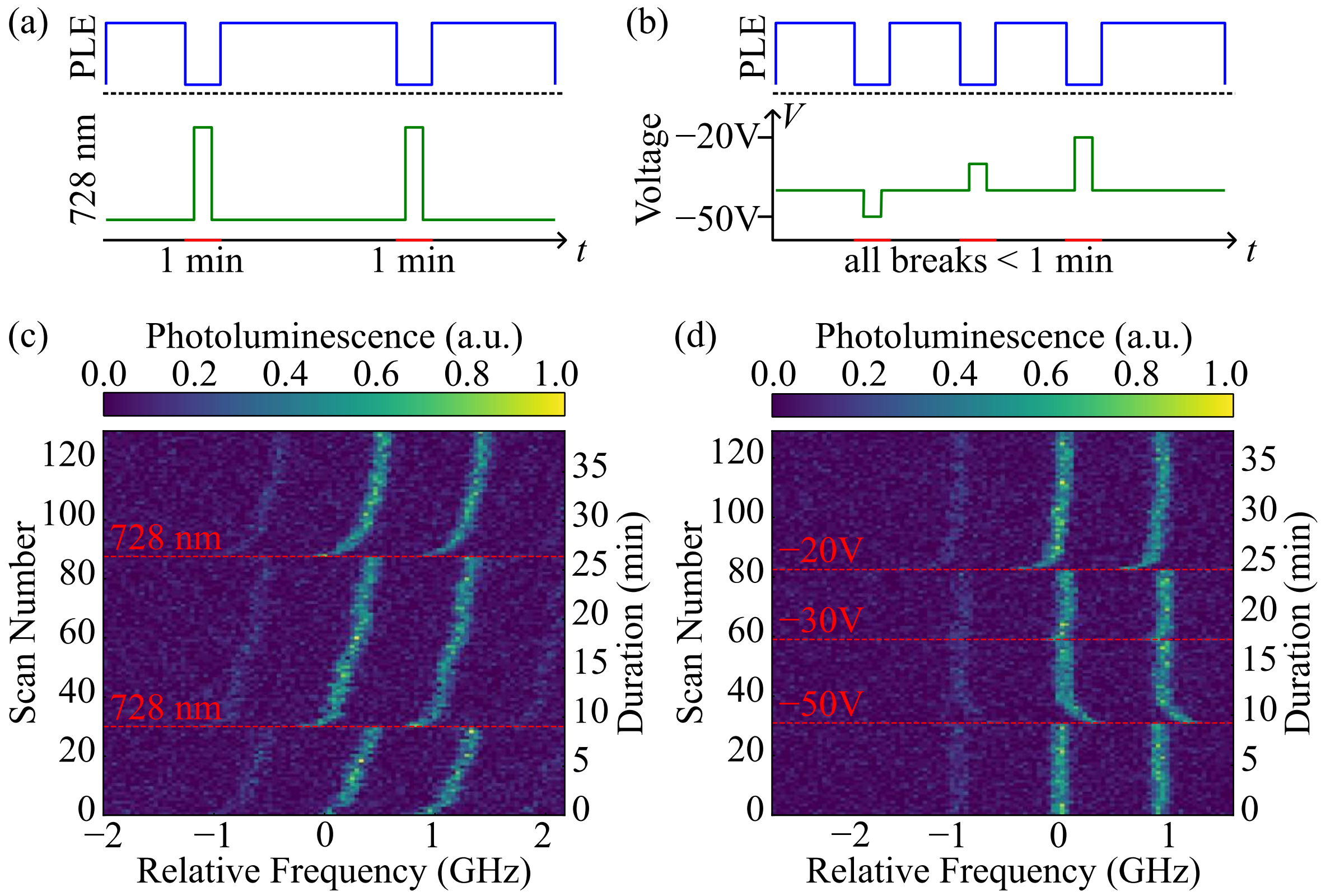

**Figure 8.** Measurement schemes (top) and corresponding time-resolved PLE series (bottom) illustrating transient shifts in the resonance frequency. In all plots, the red horizontal lines indicate specific time intervals where the measurement was intentionally interrupted to apply a perturbation; the scan was then continued under the identical external conditions that were present before the pause. (a) & (c) Influence of off-resonant laser irradiation, where the pauses mark extended exposure to the 728 nm laser. (b) & (d) Influence of previously applied voltages, showing the transient relaxation of the resonance frequency after temporarily changing the reverse bias during the pause and resuming the scan at a constant voltage of −40 V. The x-axis in the measurement schemes (top) is not to scale. The time-resolved PLE series (bottom) exhibit distinct side peaks due to higher-order modulation (see SI 3).

## 4. Discussion

### 4.1 Determination of the Local Electric Field in the pin-Diode

The integration of silicon vacancies within specific regions of the pin-diode enables the precise *in situ* mapping of internal electrostatics. Color center No. 1 is located in the region near the junction where severe electric field crowding occurs as shown in Figure 5. Because this area of high local electric field strength is also where the device typically experiences electrical breakdown, utilizing color centers in this specific region provides a unique possibility to directly monitor and study critical breakdown phenomena at the microscopic level.

However, comparing the local electric fields extracted from the optical PLE measurements (Figure 6) with those predicted by the TCAD simulations (Figure 4) reveals a notable discrepancy. Although the room-temperature forward *I-V* calibration showed strong macroscopic agreement, the simulated reverse-bias field gradients diverge substantially from the cryogenic PLE measurements: Specifically, the experimentally extracted electric field at color center No. 1 is 224% higher than simulated, whereas at color center No. 2 it is 41% lower. While the macroscopic depletion electrostatics of the slightly doped intrinsic region are often approximated as temperature-independent [2], some discrepancy can be explained by uncertainties in the exact three-dimensional position (particularly depth) of the color centers, as well as cryogenic dopant freeze-out at the implanted p-i junction and interface charge trapping at 4 K. Ultimately, this highlights the potential of using integrated color centers as quantum sensors to probe device physics at cryogenic temperatures, where established TCAD transport models and parameters remain scarce.

Resonant ODMR was used to investigate ground-state spin properties. Under an applied reverse bias, the observed Stark shift of the ODMR resonance frequency is larger than can be explained by the axial electric field (derived from the PLE measurements), using the coupling parameters published in Ref. [23]. The theoretical ODMR frequency under an electric field is described by the ground state spin Hamiltonian, mathematically expressed as follows:

$$f_{\mathrm{ODMR}}(E) = 2\sqrt{\left(D + d_{\parallel}^{GS} E_{\parallel}\right)^2 + \left(d_{\perp}^{GS} E_{\perp}\right)^2} \tag{2}$$

Using the coupling parameters established in Ref. [23] ($d_{\parallel}^{GS} = -15$ MHz/(MV/cm) and $d_{\perp}^{GS} = 16$ MHz/(MV/cm)) the expected ODMR shift based purely on the axial field ($E_{\parallel}$) was determined to be 229 kHz, thus underestimating the experimentally observed shift (313 ± 66 kHz – Figure 7). Furthermore, according to Equation 2, the presence of any perpendicular electric field components ($E_{\perp}$) would mathematically dampen the axial shift due to the quadratic addition under the square root. Consequently, this enhanced spin-electric shift cannot be reconciled solely using literature coupling constants, pointing toward either local strain-electric interplay near the p-i junction or a temperature- and doping-dependent modification of the ground-state spin-electric susceptibility.

### 4.2 Effects on Resonance Frequency

The observed transient shift of the resonance frequency following an external perturbation can be accurately modeled using an exponential decay function, where the characteristic relaxation lifetime $\tau$ serves as the primary fitting parameter (as exemplified in Figure 9a). The observed relaxation lifetime was determined to be 187 s for the third (topmost) dataset shown in Figure 8c.

This transient behavior could be explained by a localized redistribution of charge carriers. The application of the high-energy, off-resonant initialization laser can ionize nearby defects located either at the oxide interface or within the epitaxial layer itself [2, 4, 21]. Once ionized, the mobile charge carriers are swept away by the macroscopic electric field of the reverse-biased diode, leaving behind localized, fixed charges. These remaining fixed charges act as sinks for the electric field lines, effectively reducing the net electric field experienced locally by the $V_2$ silicon vacancy and thus shifting its optical transition frequency via the Stark effect. Over time, mobile charge carriers – such as those from the diode's reverse current – gradually fall back into these empty traps, restoring the local electric field to its original magnitude. Because the intrinsic region is depleted of free carriers at cryogenic temperatures, this refilling process might be rate-limited, resulting in long time constants required to reach equilibrium.

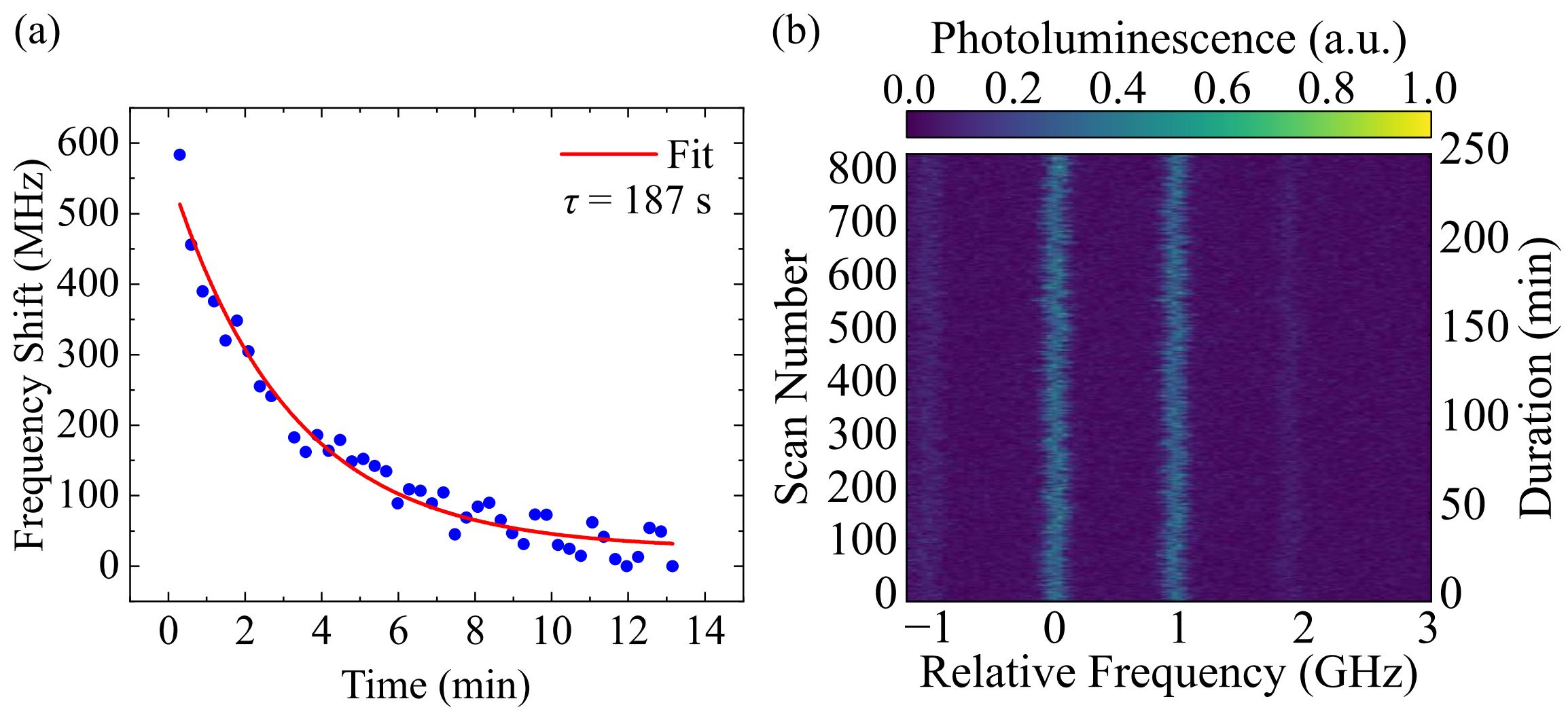

**Figure 9.** (a) Transient frequency shift over time after the removal of the off-resonant laser perturbation, fitted with an exponential decay function. (b) Long-term time-resolved PLE series over a duration of 4 hours at a constant reverse bias of −40 V, demonstrating high spectral stability. The time-resolved PLE series exhibit distinct side peaks due to higher-order modulation (see SI 3).

Analogously, the dependence on previously applied voltages observed in Figure 8d can be attributed to slow charge trapping and de-trapping dynamics at deep defect levels or at the $SiO_2$/4H-SiC interface following a change in the depletion boundary. Stepping the reverse bias temporarily alters the quasi-Fermi levels and the spatial extent of the space-charge region, modifying the local trap occupancy; upon returning to −40 V, the non-equilibrium fixed charge distribution temporarily screens (after a step from −50 V) or enhances (after a step from −20 V or −30 V) the local electric field until thermal or leakage-assisted relaxation restores steady-state equilibrium.

The influence of the laser and the resulting transient nature of these PLE measurements provide an important perspective on the spectral instabilities frequently discussed in literature, such as spectral wandering or spectral jumping. We hypothesize that these phenomena share the exact same underlying physical mechanism involving localized charge fluctuations. Depending on the measurement speed and integration time relative to the charge trapping and de-trapping time constants, this single underlying mechanism manifests either as discrete spectral jumps (when $\tau$ is short relative to the scan acquisition time) or as continuous spectral drift and wandering (when $\tau$ is long).

Crucially, once this slow charge redistribution process concludes and the local electrostatic environment reaches equilibrium, the resonance frequency exhibits exceptional long-term stability. As demonstrated in Figure 9b, a highly stable emission without significant spectral drifting can be observed over continuous measurement periods spanning several hours.

### 4.3 Spectral Matching of Distinct $V_2$ Emitters

The generation of indistinguishable photons is a critical requirement for realizing quantum entanglement and executing advanced quantum computing protocols. However, the specific local environment surrounding each defect, such as its spatial distance to dopant atoms, inherently causes slight variations in the ZPL position of individual color centers as given in Section 3.2. Left uncompensated, this inherent spectral variation prevents the generation of indistinguishable photons required for multi-qubit optical interference [26].

The integration of lateral pin-diodes provides an effective solution to this challenge through the application of the DC Stark effect [2]. By individually tuning the local electric field via the applied reverse bias, the optical transition frequencies of different emitters can be precisely controlled and matched in their mean emission frequency, as shown in Figure 10. Both the $A_1$ as well as $A_2$ transition frequencies of two distinct $V_2$ silicon vacancies can be successfully overlapped by applying specific voltages to compensate for their local environmental differences.

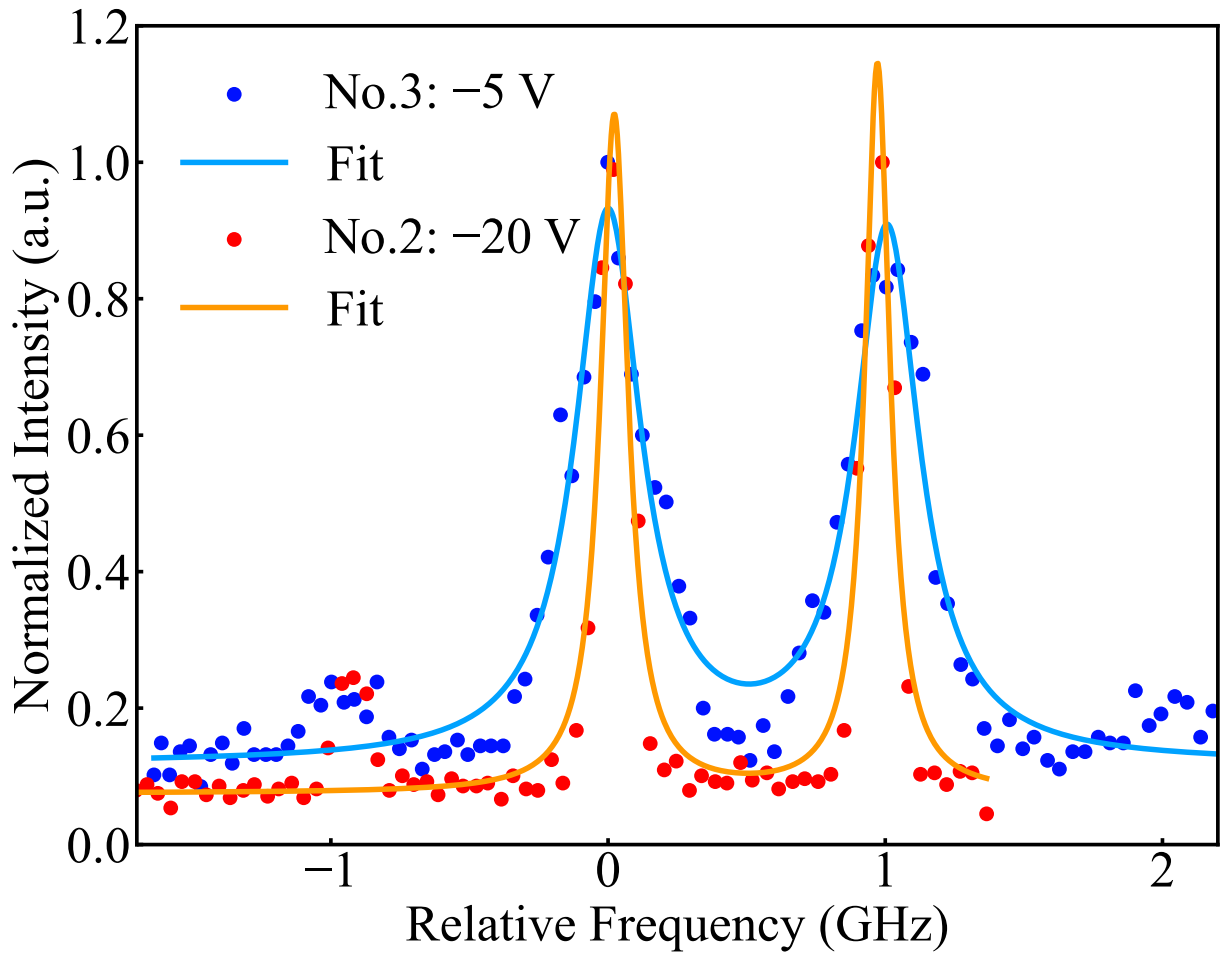

**Figure 10.** Overlapping PLE spectra of two distinct $V_2$ silicon vacancies (color center No. 3 at −5 V and No. 2 at −20 V). The externally applied electric field compensates for local environmental variations, successfully overlapping the optical transitions of both emitters.

Once the transient charging effects discussed in Section 4.2 have settled and the local electrostatic environment reaches equilibrium, these tuned spectral lines show excellent long-term stability. While the PLE measurements presented here probe the transitions via phonon-sideband detection, this precise tunability and long-term stability – combined with the identical c-axis dipole polarization of both transitions – fulfill the core spectral requirement for interfering coherently emitted ZPL photons. This provides a direct testbed for two-photon Hong-Ou-Mandel interference at a fixed diode bias as shown in Ref. [38].

Furthermore, the narrow optical linewidths of 69.4 MHz and 75.6 MHz, shown in the inset in Figure 6, achieved in these fully processed devices are highly competitive with those reported in pristine, unprocessed c-plane samples [2, 4, 12, 39]. This finding validates the high crystal quality and low background doping of the epitaxial a-plane layer, demonstrating that these pristine quantum properties are successfully maintained despite subjecting the wafer to a full, scalable CMOS processing sequence. Consequently, this confirms the viability of a-plane 4H-SiC as a robust and scalable platform for integrating high-performance quantum emitters alongside classical CMOS control electronics.

## 5. Conclusion

In this work, $V_2$ silicon vacancies were successfully integrated into lateral pin-diodes on 4H-SiC a-plane wafers using a scalable CMOS process. The a-plane orientation enabled out-of-plane resonant optical excitation, while the narrow optical linewidths of 69.4 MHz and 75.6 MHz confirmed the preservation of high crystal quality despite the extensive fabrication steps. By utilizing these color centers as embedded quantum sensors, local electric fields via the Stark effect were determined. The measurements revealed significant discrepancies with macroscopic TCAD simulations, particularly in regions of field crowding, demonstrating the unique utility of embedded spin defects for probing semiconductor device physics at cryogenic temperatures.

Furthermore, the precise spectral control required for quantum computing applications was demonstrated. By applying a reverse bias, the resulting Stark shift successfully compensated for the natural variations in the local environment of individual defects. Following initial transient charge fluctuations, the emitters exhibited excellent long-term spectral stability, allowing the resonance lines of separate vacancies to be overlapped. This capability to deterministically match the emission frequencies of distinct defects confirms the viability of a-plane 4H-SiC as a robust platform for scalable quantum photonic integrated circuits and future interference experiments.

## CRediT Author Statement

**Jannik H. Schwarberg:** Conceptualization, Methodology, Investigation, Validation, Formal Analysis, Visualization, Writing – Original Draft; **Fedor Hrunski:** Investigation, Validation,

Writing – Review and Editing; **Maximilian Hollendonner:** Investigation, Validation, Writing – Review and Editing; **Wolfgang Knolle:** Investigation; **Mathias Rommel:** Supervision, Writing – Review and Editing; **Fabian J. Magerl:** Writing – Review and Editing; **Jan F. Dick:** Writing – Review and Editing; **Jörg Schulze:** Supervision

## Acknowledgements

We would like to thank Aixtron and Birgit Kallinger for carrying out the epitaxy processes. Special thanks to the cleanroom team of LEB and IISB for their support in the fabrication of the samples. Moreover, we would like to thank Georg Kruse and Christophe Pixius for providing the in-house Python tool for the TCAD calibration. Lastly, we would like to thank Roland Nagy for the provision of measuring equipment and the fruitful discussions of the optical measurements and the results observed. The authors acknowledge the use of the generative AI tool Gemini (Google) during the preparation of this manuscript. The tool was utilized to assist in translating and adapting excerpts from a German doctoral thesis into English to help draft the manuscript text. The authors thoroughly reviewed, edited, and verified all AI-assisted content and take full responsibility for the accuracy, originality, and integrity of the final published work. Funded by the Deutsche Forschungsgemeinschaft (DFG, German Research Foundation) – 532769051. The authors declare no competing interests.

## Data availability statement

The data that support the findings of this study are available from the corresponding author upon reasonable request.

# Supplementary Information – Electrical Stark Tuning and In Situ Field Sensing of $V_2$ Silicon Vacancies in a 4H-SiC a-Plane Lateral pin-Diode

**Jannik H. Schwarberg[1,*], Fedor D. Hrunski[2], Maximilian Hollendonner[2], Wolfgang Knolle[3], Mathias Rommel[4], Fabian J. Magerl[1], Jan F. Dick[1,4] and Jörg Schulze[1,4]**

[1] Chair of Electron Devices, Friedrich-Alexander-Universität Erlangen-Nürnberg, 91058 Erlangen, Germany
[2] Institute of Applied Quantum Technologies, Friedrich-Alexander-Universität Erlangen-Nürnberg, 91058 Erlangen, Germany
[3] Leibniz-Institut für Oberflächenmodifizierung, 04318 Leipzig, Germany
[4] Fraunhofer Institute for Integrated Systems and Device Technology IISB, 91058 Erlangen, Germany

*Author to whom any correspondence should be addressed.

**E-mail:** jannik.schwarberg@fau.de

## SI 1 Details of Device Manufacturing

The devices were fabricated using a scalable, CMOS process [1, 2] on 35 mm on-axis 4H-SiC a-plane substrates featuring a 10 μm-thick epitaxial layer with a low n-type doping concentration of $2.4 \times 10^{13}$ $cm^{-3}$ [3].

To form the highly doped p-type and n-type regions of the pin-diodes following the fabrication sequence detailed in Ref. [2], aluminum and nitrogen ions were implanted through a 30 nm scattering oxide. The dopants were subsequently activated by thermal annealing at 1700 °C in an argon atmosphere with a carbon capping layer. An initial 50 nm-thick thermal oxide – which serves as the gate dielectric in the standard CMOS flow – was grown by dry thermal oxidation at 1300 °C for 6.5 minutes [2], followed by a 30-minute anneal in nitric oxide. After deposition of an insulation oxide, the ohmic contacts were defined using a self-aligned 70 nm NiAl layer and silicided by rapid thermal processing at 980 °C for 2 minutes. Device metallization consisted of a 40 nm Ti / 400 nm Pt / 40 nm Ti stack, and the devices were fully passivated using a $SiO_2/Si_3N_4/SiO_2$ layer stack.

To enable resonant laser excitation of the defect centers, an optical window with low noise surface passivation was structured above the intrinsic region as established in Ref. [4]. During standard CMOS processing, a 500 nm n-type doped polycrystalline silicon (poly-Si) layer (typically serving as the gate electrode) was deposited to protect the underlying 50 nm thermal oxide. After completing the high-aspect-ratio dry etching steps (slight over-etching for wafer conformity) for the insulation and passivation layers, the remaining poly-Si protective layer (remaining thickness < 350 nm) was removed using a two-step wet chemical etch: a 240 s dip in 1% HF followed by a highly selective self-limiting etch using 5% TMAH at 70 °C. This approach successfully left the 50 nm thermal oxide intact.

Following full device processing, silicon vacancies were generated via electron irradiation (energy of ~ 4.5 MeV, dose of $3 \times 10^{12}$ $cm^{-2}$), followed by a thermal anneal at 600 °C in vacuum for 30 minutes. The schematic of the final device is shown in figure 1 of the main text. The $V_{Si}$ are only shown in the intrinsic (lowly n-type doped) region, since they are only observable in low n-type doping due to their charge state [5, 6].

## SI 2 Models used for TCAD Simulations

Two-dimensional numerical device simulations were conducted using Synopsys Sentaurus TCAD (version W-2024.09-SP1) to evaluate the internal electrostatics and electric field profiles within the lateral pin-diodes.

The physical structure was constructed within the sprocess module to accurately reproduce the fabricated lateral diode geometry. The device coordinate system was defined with the x-axis aligned parallel to the crystallographic c-axis. Doping profiles of the implanted p-type and n-type regions were introduced using Monte Carlo ion implantation simulations, and substrate consumption during thermal oxidation was accounted for. The final device in Sentaurus TCAD is shown in figure 2 of the main text.

For the physical description at 300 K, the following key electrical models were utilized: Charge carrier mobility was modeled both in terms of its doping dependence (Arora model) and its saturation at high electric field strengths (high-field saturation). The recombination mechanisms included Shockley-Read-Hall and Auger recombination. The crystallographic anisotropy of the material was accounted for by using anisotropic models for mobility and the parameters in Poisson's equation. In addition, the incomplete ionization model for the dopants was activated. For the numerical solution of the system of equations, the SUPER solver (method: Blocked) was employed.

To accurately calibrate the forward characteristics to the measured $I$-$V$ curves prior to simulating reverse-bias field distributions, the p-type contact was modeled as an imperfect ohmic contact exhibiting Schottky barrier behavior [2], using a non-local mesh hole-barrier tunneling model with barrier lowering. A parameter calibration routine was executed using an in-house Python tool employing numerical optimization algorithms. The optimization calibrated five partly interdependent physical and geometric parameters against experimental forward $I$-$V$ characteristics: effective carrier lifetime, p-contact Schottky barrier height, effective hole tunneling mass, external parasitic series resistance, and the effective out-of-plane conduction width.

## SI 3 Optical Measurement Setup

The optical measurements were performed using a home-built confocal microscope setup based primarily on free-space optics [7], specifically designed for the investigation of $V_2$ silicon vacancies. For resonant excitation, a tunable diode laser at 916 nm (Toptica DL Pro) with laser power set to 20 nW was utilized, while off-resonant excitation was achieved using a 728 nm laser (Toptica iBeam smart). The excitation of the sample and the collection of the emitted photons were conducted through a high numerical aperture objective (Attocube LT-APO). The drive frequencies for the microwave signal were generated using an Arbitrary Waveform Generator (AWG, Windfreak SynthNV PRO). The same AWG was used to drive the electro-optical modulator (EOM, Exail MPX-950-LN-10) used for the double tone PLE measurements. While the primary measurement utilizes the $n = 1$ modulation, the periodic frequency modulation via the electro-optical modulator also generates higher-order sidebands. Consequently, the resulting PLE spectra exhibit distinct $n = 2$ side peaks that appear significantly weaker in intensity compared to the main signal. To filter the photons of the resonant excitation laser from the phonon-sideband emission of the $V_2$, a 950 nm longpass filter (Thorlabs FELH0950) was employed. The sideband emission was then coupled into optical fibers and directed to a superconducting nanowire single-photon detector (SNSPD, Single Quantum), with the signals recorded by a time tagger (Swabian Instruments Time Tagger Ultra). During all cryogenic measurements, the sample was housed inside a cryostat (Qinu) operated at a temperature of 4 K.

## SI 4 Comment on Local Lorentz Field Approximation

To compare the simulated macroscopic electric field in c-direction $E_{||}$ inside the pin-diode with the microscopic field $E_{\mathrm{loc},||}$ acting directly on the $V_2$ silicon vacancy, one can apply the Lorentz local field approximation, which accounts for the local polarization of the surrounding host lattice according to Ref. [8]:

$$E_{\mathrm{loc},||} = \frac{\varepsilon_r + 2}{3} E_{||} \tag{1}$$

Given our lateral device geometry on a-plane 4H-SiC with the electric field oriented parallel to the crystallographic $c$-axis ($E \parallel c$), using the corresponding relative permittivity $\varepsilon_r = 10.03$ yields $E_{\mathrm{loc},||} \approx 4E_{||}$. To make comparison with the simulated electric field easier, we decided to consequently use the macroscopic field throughout the main text. Following the Stark shift description used in Ref. [9], the frequency shift $\Delta f_{PLE}$ of the zero-phonon line transition is modeled up to second order:

$$\Delta f_{PLE} = d_{||} \cdot E_{||} - \frac{1}{2} a_{||} \cdot {E_{||}}^2 \tag{2}$$

Here $d_{||}$ and $a_{||}$ denote the linear (dipole moment change) and quadratic (polarizability change) coupling parameters, respectively. When translating coupling parameters derived under the local Lorentz field approximation to the macroscopic electric field (or vice versa), substituting $E_{\mathrm{loc},||} \approx 4E_{||}$ requires scaling the linear coefficient by a factor of 4 ($d_{||} \rightarrow 4d_{||}$) and – due to the squared electric field dependence in the second-order term – scaling the quadratic coefficient by a factor of 16 ($a_{||} \rightarrow 16a_{||}$).